# Quantum non-locality, causality and mistrustful cryptography

*Muhammad Nadeem*
*Department of Basic Sciences,*
*School of Electrical Engineering and Computer Science*
*National University of Sciences and Technology (NUST)*
*H-12 Islamabad, Pakistan*
*muhammad.nadeem@seecs.edu.pk*

Here we propose a general relativistic quantum framework for cryptography that exploits the fascinating connection of quantum non-locality and special theory of relativity with cryptography. The underlying principle of unconditional security is causality and two-fold quantum non-local correlations: first entanglement swapping and then teleportation over causally independent entangled systems. We show that the proposed framework has following remarkable and novel features in mistrustful cryptography: (i) It helps in defining a new notion of oblivious transfer where both the data transferred and the transfer position remains oblivious. (ii) The authenticity and integrity of the data transferred is guaranteed by the fundamental principles of quantum theory instead of computational complexity. (iii) It directly leads to unconditionally secure and deterministic two-sided two-party computation which is currently considered to be impossible. (iv) the proposed framework turns out to be asynchronous ideal coin tossing with zero bias which has not been achieved previously. (v) The same framework also implies unconditionally secure bit commitment. Finally, the combination of quantum non-locality and theory of relativity as discussed here can easily be generalized to multiparty setting that could be used to solve other mistrustful cryptographic tasks such as secret sharing and key agreement securely.

In the last few years, researchers have shown great excitement in the area of relativistic quantum cryptography[1-9,10-28] where causal structure of Minkowski space time or impossibility of superluminal signaling gives power to relativistic quantum cryptography in defining tasks that are not possible in non-relativistic setting, especially in mistrustful cryptography. These interesting developments give further hope for defining a more general framework in relativistic quantum theory that would be sufficient to solve all the cryptographic tasks securely.

Kilian showed that classical oblivious transfer[29-32] (OT) is a basic building block for many mistrustful cryptographic protocols, for example, two-party secure computations[33]. However, since computationally hard classical protocols can be broken, various protocols for OT based on non-relativistic[34] and relativistic quantum theory[22] have also been proposed. In existing non-relativistic quantum OT protocols, only data remains oblivious while sender/receiver can be well aware of transfer position. On the other hand, in relativistic OT protocol[22], the data can be completely determined by sender/receiver while they remain oblivious about the transfer position.

Moreover, in all previously proposed OT protocols, receiver cannot be certain that the data he received has not been altered during the protocol. Hence, currently it is known that 1-out-of-2 oblivious transfer and deterministic two-sided two-party secure computations (TPSC) are impossible in classical/non-relativistic quantum cryptography[35,36]. These impossibility results have also been extended to relativistic quantum cryptography[37]. However, relativistic quantum cryptography gives hope for secure implementation of nondeterministic two-sided TPSC and hence variable-bias coin tossing[38]. Moreover, asynchronous ideal coin tossing is impossible in classical/non-relativistic quantum cryptography[39] while only synchronous ideal coin tossing is possible if impossibility of superluminal signaling is considered[40].

Furthermore, bit commitment is another very important and basic cryptographic protocol that is impossible in classical/nor-relativistic quantum cryptography[41-43] but has been proved to be possible in relativistic quantum theory[10,24,26]. These no-go theorems show the limitations of classical/non-relativistic quantum cryptography while possibility results show that relativity adds its weight, and hence gives more power, towards quantum cryptography to evade such no-go theorems.

At this point, we would like to discuss an important quantum mechanical concept, non-locality, which has an interesting connection with cryptography and cryptanalysis. Non-local Einstein-Podolsky-Rosen (EPR) type correlations[44] solves the very basic ingredient of cryptography, QKD[45], that gives unconditionally secure means for secret communications between distant parties. On the other hand, in mistrustful cryptography, a dishonest party can exploit the non-locality (EPR types quantum attacks) to cheat successfully[35,41-43].

In this work, we exploit the fascinating connections of quantum non-locality and relativity with cryptography and show that the combination of relativity with non-locality favors cryptography rather than cryptanalysis. We propose a general relativistic quantum framework for cryptography and show that the proposed framework proves to be a building block for many interesting mistrustful cryptographic protocols that are considered to be impossible. For example, it directly leads to (i) a new notion of OT where both the data transferred and the transfer position remain oblivious, (ii) deterministic two-sided TPSC, (iii) asynchronous ideal coin tossing with zero bias, and (iv) unconditionally secure bit commitment.

In fact, the proposed framework could be used to solve other mistrustful cryptographic tasks with guarantee of authenticity and integrity of the data transferred along with unconditional security against Mayers and Lo-Chau (MLC) attacks[35,39, 41-43] and quantum attacks based on non-local instantaneous computations[46]. The proposed framework attains unconditional security through combination of causality and two-fold quantum non-local correlations: first entanglement swapping[47] and then teleportation[48] over causally independent entangled systems.

## Mistrustful Cryptography

In a standard mistrustful cryptography, communicating parties do not trust each other and security is concerned against these parties only (internal eavesdropping); everything outside secure laboratories of the communicating parties is assumed to be insecure. This subject is very divers and has gained a lot of attention especially in quantum setup; however, we focus here only those cryptographic primitives which could be used as a building block for implementing many other protocols. For example, (i) OT and (hence) two-sided TPSC and ideal coin tossing and (ii) unconditionally secure bit commitment.

**Oblivious transfer:** OT was originally defined by Rabin where sender Alice sends a 1-bit message to the receiver Bob who can only receive the message with probability no more than half[29]. The security of the protocol relies on the fact that Bob can find out whether or not he got the 1-bit message from Alice after the completion of protocol but Alice remains oblivious about it. In a related notion, 1-out-of-2 OT, Alice sends two 1-bit messages to Bob who can only receive one of them and remains ignorant about the other while Alice remains entirely oblivious to which of the two messages Bob received[30,31]. It is shown later by Crépeau that both of these notions of OT are equivalent[32].

**Two-sided two-party secure computation:** Two-sided TPSC enables two distant parties Alice and Bob to compute a function $f(a,b)$ where $a$ and $b$ are inputs from Alice and Bob respectively. The protocol is said to be secure if it fulfils following security requirements: (i) both Alice and Bob learn output of $f(a,b)$ deterministically. (ii) Alice learns nothing about Bob's input $b$ and (iii) Bob learns nothing about Alice's input $a$. The no-go theorems for secure two-party computations are based on possibilities that one party, say Bob, can also compute $f(a,b')$ where $b' \in \{b_1,b_2,\ldots\}$. That is, Bob can cheat by computing the value of the function $f$ for all of his inputs $b'$ and hence violate the security requirement of single input from each party. Lo[35] has shown that Bob can do this by applying unitary transformations on his own quantum system $\mathcal{H}_B$. That is, the system $\mathcal{H}_B$ kept by Bob must be an eigenstate of the measurement operator that he uses for computing $f(a,b)$. Being an eigenstate, $\mathcal{H}_B$ remains undisturbed by Bob's measurement that makes computation of $f(a,b')$ feasible.

**Ideal quantum coin tossing:** Coin tossing[49] is another fundamental primitive function in communication that allows distant mistrustful parties Alice and Bob to agree on a random data. Coin tossing is said to be ideal if it follows: (i). It results in three possible outcomes $\gamma$: $\gamma_+ = +$, $\gamma_- = -$ or $\gamma_\pm$ = invalid. (ii). Outcome $\gamma_+$ and $\gamma_-$ occurs with equal probability $P_+ = P_- = 1/2$ and both parties have equal cheating probabilities, $P_\gamma^A = P_\gamma^B = P_\gamma$, which means that the coin tossing is fair. (iii). If both parties are honest, the outcome $\gamma_\pm$ = invalid never occurs; $P_\pm = 0$. (iv). If any one of the parties is dishonest, the outcome invalid occurs with probability $P_\pm = 1$.

**Bit commitment:** A bit commitment is also an important cryptographic scheme between two mistrustful parties, committer (Alice) and receiver (Bob), where Alice commits herself to a specific bit $a$ in the commitment phase. In this phase or during the scheme, Bob should not be able to extract the bit value. In the revealing phase, however, it must be possible for Bob to know the genuine bit value $a$ with absolute guarantee when Alice reveals the committed bit and Alice should not be able to change her mind about the value of the bit $a$.

## Proposed framework for relativistic quantum cryptography

In a relativistic quantum cryptographic setup proposed by Adrian Kent, background space time is approximately Minkowski and communicating parties Alice and Bob are not the individuals but are agencies having distributed agents throughout the space time. The agencies are assumed to possess secure sites in a given inertial frame and can communicate with each other by sending quantum/classical signals at near light speed c=1. Moreover, the agencies have unlimited powers of information processing and efficient technology (quantum computers) and are restricted from cheating by principals of quantum theory only. Alice and Bob can communicate with their respective agents securely; however, all the quantum/classical channels between Alice and Bob are insecure. Both agencies have powers of instantaneous computation and time for information processing at their secure sites is assumed to be negligibly small. If one of the agencies sends a quantum/classical signal from point $(x,0)$, then after some fixed time t > 0, the light-like separated agents from the sender in some given inertial frame can receive the signal on a special sphere of radius t and centered at $x$.

In the proposed framework for relativistic quantum cryptography we assume that Alice has two agents $A_1$ and $A_2$ while Bob has three agents B', $B_1$ and $B_2$ respectively. Suppose Alice possesses secure site at $(x_a, t_a)$, space-like separated from Bob and his agent B' at positions $(x_b, t_b)$ and $(x_{b'}, t_{b'})$ respectively such that $t_a = t_b = t_{b'} = t$ , $x_b < x_a < x_{b'}$ and $x_a = (x_{b'} - x_b)/2$ . However, Bob's agents $B_1$ and $B_2$ are at arbitrary positions $(x_{b_1}, t_{b_1})$ and $(x_{b_2}, t_{b_2})$ in the causal future of Alice respectively. Alice only knows the directions where $B_1$ and $B_2$ can receive the data but not their exact positions; they are light-like separated from Alice. Similarly, Alice's agents $A_1$ and $A_2$ are light-like separated from Bob in his causal future at positions $(x_{a_1}, t_{a_1})$ and $(x_{a_2}, t_{a_2})$ unknown to Bob.

Alice, Bob and his agent B' share quantum system $\mathcal{H}_S = \mathcal{H} \otimes \mathcal{H}_{AB} \otimes \mathcal{H}_{A'B'}$ where $\mathcal{H} = |\varphi\rangle$ is kept by Bob while entangled systems $\mathcal{H}_{AB} = \mathcal{H}_A \otimes \mathcal{H}_B$ and $\mathcal{H}_{A'B'} = \mathcal{H}_{A'} \otimes \mathcal{H}_{B'}$ are shared with Alice by Bob and B' respectively. Alice and Bob give input data $u_a u_{a'} \in \{00,01,10,11\}$ and $u_b u_{b'} \in \{00,01,10,11\}$ to the shared quantum system $\mathcal{H}_S$ by applying unitary transformations $U^{u_a} U^{u_{a'}}$ and $U^{u_b} U^{u_{b'}}$ on $\mathcal{H}_S$ respectively as follows: Alice applies transformations $U^{u_a} U^{u_{a'}}$ on $\mathcal{H}_{A'}$, performs Bell state measurement[50] (BSM) on $\mathcal{H}_A \otimes \mathcal{H}_{A'}$, and publically announces her BSM result $\alpha\alpha'$ . As a result, Bob and his agent B' get entangled; $\mathcal{H}_{BB'} = \mathcal{H}_B \otimes \mathcal{H}_{B'}$. Simultaneously, Bob applies transformations $U^{u_b} U^{u_{b'}}$ on $\mathcal{H} = |\varphi\rangle$ , teleports the quantum state $U^{u_b} U^{u_{b'}} |\varphi\rangle$ to B' by applying local Bell operator $\beta_b$ on $\mathcal{H} \otimes \mathcal{H}_B$, and publically announces his BSM result $\beta\beta'$. He also sends $U^{u_b} U^{u_{b'}} |\varphi\rangle$ to either $A_1$ or $A_2$. Instantly B' (measures if required and) sends his system $\mathcal{H}_{B'} = U^i U^{u_b} U^{u_{b'}} |\varphi\rangle$ to Alice where $U^i$ is teleportation encoding. Alice applies further unitary transformations $U^{u_a} U^{u_{a'}}$ and sends $U^{u_a} U^{u_{a'}} U^i U^{u_b} U^{u_{b'}} |\varphi\rangle$ to either $B_1$ or $B_2$.

Alice and Bob validate the protocol if non-local quantum correlations between Alice's transformation $U^{u_a} U^{u_{a'}}$ , Bob's transformation $U^{u_b} U^{u_{b'}}$ , and teleportation encoding $U^i$ are consistent with BSM results ($\alpha\alpha'$ and $\beta\beta'$) of Alice and Bob on initially shared quantum system $\mathcal{H}_S$. This framework is secure in general and against Alice/Bob in particular through quantum non-local correlations and causal structure of Minkowski space time.

Here $\mathcal{H}$, $\mathcal{H}_{AB}$ and $\mathcal{H}_{A'B'}$ can be any higher dimensional quantum systems where Alice and Bob can execute any set of suitable unitary transformations $U^{u_a} U^{u_{a'}}$ and $U^{u_b} U^{u_{b'}}$ respectively. However, to make the analysis simple, we assume in the rest of the discussion that $\mathcal{H} = |\varphi\rangle = |\pm\rangle$ where $|\pm\rangle = (|0\rangle \pm |1\rangle)/\sqrt{2}$ while both $\mathcal{H}_A \otimes \mathcal{H}_B = (C^2) \otimes (C^2)$ and $\mathcal{H}_{A'} \otimes \mathcal{H}_{B'} = (C^2) \otimes (C^2)$ are 2-qubit maximally entangled systems with Bell basis

$$|u_m u_n\rangle = \frac{|0\rangle |u_n\rangle + (-1)^{u_m} |1\rangle |1 \oplus u_n\rangle}{\sqrt{2}} \quad (1)$$

where $u_m$ and $u_n \in \{0,1\}$ and $\oplus$ denotes addition with mod 2. The unitary transformations used by Alice and Bob will be $U^{u_a} U^{u_{a'}} = \sigma_z^{u_a} \sigma_x^{u_{a'}}$ and $U^{u_b} U^{u_{b'}} = \sigma_z^{u_b} \sigma_x^{u_{b'}}$ respectively where they agree on a code: if sender S (Alice/Bob) applies unitary transformation $I$, $\sigma_x$, $\sigma_z$, or $\sigma_z \sigma_x$ on the quantum state $|\varphi\rangle \in \mathcal{H}_S$, he/she is actually giving input data 00, 01, 10 or 11 to the system $\mathcal{H}_S$ respectively. Since $\sigma_z^{u_s} \sigma_x^{u_{s'}} |\varphi\rangle = \sigma_z^{u_s} \sigma_x^{1 \oplus u_{s'}} |\varphi\rangle$ if $|\varphi\rangle = |\pm\rangle$ , where we ignore the overall phase factor, we restate the code as follows: to send data $u_s u_{s'} \in \{00,01\}$ or $u_s u_{s'} \in \{10,11\}$ , sender S (Alice/Bob) applies corresponding Pauli transformations $\sigma_s \in \{\sigma_z^{u_s} \sigma_x^{u_{s'}}, \sigma_z^{u_s} \sigma_x^{1 \oplus u_{s'}}\}$ on the quantum state $|\varphi\rangle \in \mathcal{H}_S$. With two dimensional quantum registers, explicit procedure for our proposed framework is described below:

(1). At t=0, Bob and his agent B' prepare EPR pairs $|u_a u_b\rangle \in \mathcal{H}_A \otimes \mathcal{H}_B$ and $|u_{a'} u_{b'}\rangle \in \mathcal{H}_{A'} \otimes \mathcal{H}_{B'}$ respectively and each sends first qubit to Alice.

(2). At $t = x_a - x_b$, Alice applies Pauli transformation $\sigma_a \in \{\sigma_z^{u_a} \sigma_x^{u_{a'}}, \sigma_z^{u_a} \sigma_x^{1 \oplus u_{a'}}\}$ on $|u_{a'}\rangle$, performs BSM on qubits $|u_a\rangle$ and $|u_{a'}\rangle$ and publically announces her BSM result $\alpha\alpha'$.

(3). At the same time $t = x_a - x_b$, Bob prepares a qubit $|\varphi\rangle = |\pm\rangle$, applies Pauli transformation $\sigma_b \in \{\sigma_z^{u_b} \sigma_x^{u_{b'}}, \sigma_z^{u_b} \sigma_x^{1 \oplus u_{b'}}\}$ and teleports the state $\sigma_b|\varphi\rangle$ to B' over EPR channel $|u_b u_{b'}\rangle \in \mathcal{H}_B \otimes \mathcal{H}_{B'}$ established between them due to BSM of Alice. As a result, $\mathcal{H}_{B'}$ becomes one of the corresponding four possible states $|\psi\rangle = \sigma_i \sigma_b |\varphi\rangle$ where teleportation encoding $\sigma_i \in \{I, \sigma_x, \sigma_z, \sigma_z \sigma_x\}$ is unknown to everyone. Simultaneously, Bob publically announces his BSM $\beta\beta'$, sends $\sigma_b|\varphi\rangle$ to either $A_1$ or $A_2$ while B' measures and sends $|\psi\rangle$ to Alice.

(4). At $t = 2(x_a - x_b)$, Alice measures $|\psi\rangle$, applies unitary transformations corresponding to her input data $\sigma_a \in \{\sigma_z^{u_a} \sigma_x^{u_{a'}}, \sigma_z^{u_a} \sigma_x^{1 \oplus u_{a'}}\}$ on $|\psi\rangle$ and immediately sends the outcome of two-sided computation $|\psi'\rangle = \sigma_a |\psi\rangle = \sigma_a \sigma_i \sigma_b |\varphi\rangle$ to either $B_1$ or $B_2$.

(5). $B_i$ receives the state $|\psi'\rangle$ from Alice, measures in $\{+,-\}$ basis and gets joint measurement outcome $\psi'$.

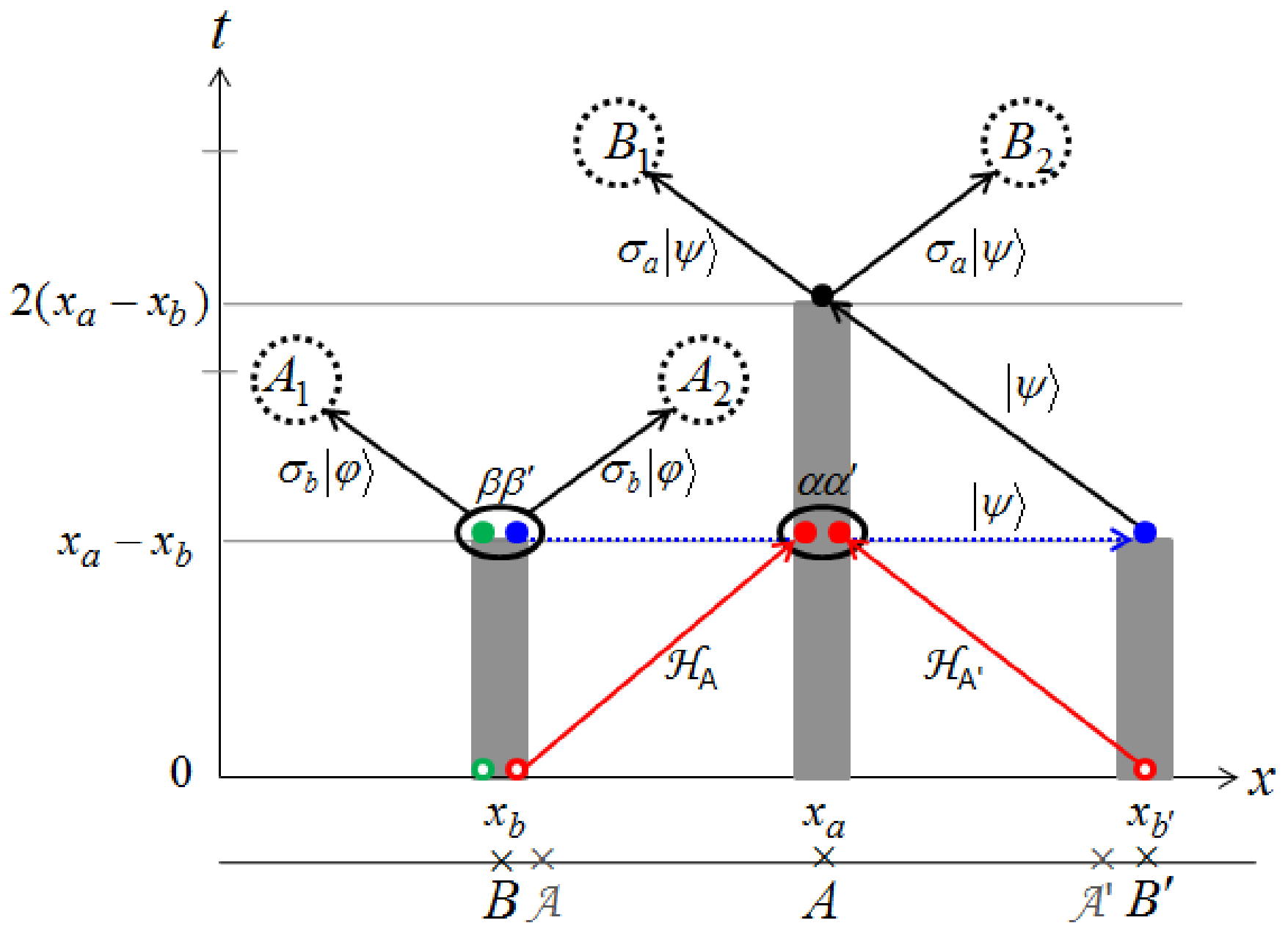

**Figure 1:** Relativistic quantum framework for cryptography in 1+1 Minkowski space time. Entangled systems $\mathcal{H}_{AB}$ and $\mathcal{H}_{A'B'}$ are represented by red particles while particle in green at Bob's site represents $\mathcal{H}=|\varphi\rangle$. At time $t = x_a - x_b$, swapped entangled system $\mathcal{H}_{BB'}$ is represented by blue color where dotted arrow shows teleportation from Bob to B'. For two-sided computations and bit commitment, Alice's agents $\mathcal{A}$ and $\mathcal{A}'$ ensure that Bob and B' are at announced space-like separated positions $x_b$ and $x_{b'}$ respectively.

For two-sided computations (TPSC & coin tossing) and bit commitment, the proposed framework strictly requires that Bob and his agent B' should have operated from their designated space-like separated positions $(x_b, t_b)$ and $(x_{b'}, t_{b'})$. Alice can insure this by assigning two more agents $\mathcal{A}$ and $\mathcal{A}'$ at disjoint sites nearby $(x_b, t_b)$ and $(x_{b'}, t_{b'})$ respectively. Instead of sending $\mathcal{H}_{\mathrm{A}}$ and $\mathcal{H}_{\mathrm{A'}}$ directly to Alice, Bob and B' will share these quantum systems with $\mathcal{A}$ and $\mathcal{A}'$ respectively who then pass these systems to Alice securely.

**Applications of proposed relativistic framework in mistrustful cryptography**

In this section, we show that the proposed framework solves the problem of OT, deterministic two-sided TPSC, asynchronous ideal coin tossing with zero bias and unconditionally secure bit commitment.

To implement OT, deterministic two-sided TPSC, and asynchronous ideal coin tossing with zero bias, let's suppose shared quantum system $\mathcal{H}_S$ is known to both Alice and Bob (priorly decided somewhere in causal past) and they use the code as follows: unitary transformation $\sigma_s \in \{I, \sigma_x\}$ correspond to classical 2-bit string $u_s u_{s'} \in \{00, 01\}$ (or bit $s = u_s \oplus u_{s'} \in \{0,1\}$ ) while those of $\sigma_s \in \{\sigma_z, \sigma_z\sigma_x\}$ correspond to classical 2-bit string $u_s u_{s'} \in \{10, 11\}$ (or bit $s = u_s \oplus u_{s'} \in \{1, 0\}$ ).

The proposed relativistic quantum framework can also be used to implement unconditionally secure bit commitment as follows: the unitary transformation $\sigma_a$ applied by Alice corresponds to her commitment $u_a u_{a'}$. By applying $\sigma_a \in \{\sigma_z^{u_a}\sigma_x^{u_{a'}}, \sigma_z^{u_a}\sigma_x^{1\oplus u_{a'}}\}$ on EPR pair $|u_{a'}u_{b'}\rangle$, Alice commits herself to the bit value $a = 0$ if $u_a = 0$ and $a = 1$ if $u_a = 1$. That is, if she wants to commit bit $a = 0$ , she applies $\sigma_a \in \{I, \sigma_x\}$ corresponding to $u_a u_{a'} \in \{00, 01\}$ and if she applies $\sigma_a \in \{\sigma_z, \sigma_z\sigma_x\}$ corresponding to $u_a u_{a'} \in \{10, 11\}$, she is committed to bit $a = 1$. In the revealing phase, Alice sends $|\psi'\rangle = \sigma_a|\psi\rangle = \sigma_a\sigma_i\sigma_b|\varphi\rangle$ where her commitment is encoded in unitary transformation $\sigma_a$.

**Oblivious Transfer:** Our proposed framework helps in defining a new notion of OT where receiver R (Alice/Bob) remains oblivious about both the data transferred and the transfer position; he/she may know both the transferred messages $\sigma_z^{u_s}\sigma_x^{u_{s'}}$ or $\sigma_z^{u_s}\sigma_x^{1\oplus u_{s'}}$ but remains oblivious about the genuine one. On the other hand, the sender S (Alice/Bob) cannot learn the transfer position even after the protocol is completed. Moreover, R accepts the data $\sigma_s$ only if he is certain that data has come from legitimate sender S, by measuring time lapse and testifying non-local quantum correlations established through local operations. Finally, in our secure OT protocol, S cannot change the data he/she started with otherwise R rejects the protocol – that is something not possible in all the previously proposed OT protocols.

**Two-sided two-party secure computation:** Our proposed framework also results in secure and deterministic two-sided TPSC of function $f(\sigma_a, \sigma_b; |\varphi\rangle)$ where $\sigma_a$ and $\sigma_b$ are unitary transformations on quantum system $\mathcal{H}_S = \mathcal{H} \otimes \mathcal{H}_{\mathrm{AB}} \otimes \mathcal{H}_{\mathrm{A'B'}}$ applied by Alice and Bob respectively. According the code, when Alice and Bob apply these transformations on $|\varphi\rangle$, they actually provide input $u_a u_{a'} \to \sigma_a$ and $u_b u_{b'} \to \sigma_b$ to the shared quantum system $\mathcal{H}_S$ respectively. At the end of the computation, both parties know the same definite outcome

$$f(\sigma_a, \sigma_b; |\varphi\rangle) = \sigma_a\sigma_i\sigma_b|\varphi\rangle \quad (2)$$

where teleportation encoding $\sigma_i$ is non-locally correlated with local operations of Alice and Bob on the shared quantum system $\mathcal{H}_S$. Bob's input $u_b u_{b'}$ remains totally random to Alice while Bob also remains oblivious about Alice's input $u_a u_{a'}$. Finally both Alice and Bob get same unbiased outcome of function $f(\sigma_a, \sigma_b; |\varphi\rangle)$ deterministically. As we show in security analysis, neither Alice nor Bob can cheat by altering their inputs to the shared quantum system $\mathcal{H}_S$ and (hence) cannot cheat to get outcome of function $f(\sigma_a, \sigma_b; |\varphi\rangle)$ as they want.

**Ideal quantum coin tossing:** Proposed framework is in fact an asynchronous ideal quantum coin tossing where both parties have equal resources and the protocol offers zero bias. That is, it fulfils all the security requirements of ideal coin tossing: $P_+ = P_- = 1/2$, zero cheating probabilities for Alice and Bob ($P_\gamma^A = P_\gamma^B = 0$), $P_\pm = 0$ if both parties are honest and $P_\pm = 1$ if any one of the parties tries cheating. Although both parties give input simultaneously at time $t = x_a - x_b$, however, proposed framework is asynchronous in the sense that Bob and Alice reveal their inputs at different times; $t = x_a - x_b$ and $t = 2(x_a - x_b)$ respectively.

**Bit commitment:** As for as bit commitment is concerned, Bob need not to send $\sigma_b|\varphi\rangle$ to $A_i$. We show in the security analysis that (i) Committer Alice cannot alter her commitment $\sigma_a$ (or $u_a u_{a'}$ or $a$) after announcement of her BSM result $\alpha\alpha'$. (ii) Receiver Bob remains unable to know Alice's committed bit $a$ during the protocol. (iii) Framework allows Bob to know the definite Alice's commitment $\sigma_a \in \{I, \sigma_x\}$ ($u_a u_{a'} \in \{00,01\}$ or $a = 0$) or $\sigma_a \in \{\sigma_z, \sigma_z\sigma_x\}$ ($u_a u_{a'} \in \{10,11\}$ or $a = 1$) at the end of protocol. (iv) Two-fold quantum non-local correlations guarantee Bob to reject the protocol if Alice tries to alter her commitment in the revealing phase.

Proposed framework can be modified for computational basis with transformations $\sigma_s \in \{\sigma_z^{u_s}\sigma_x^{u_{s'}}, \sigma_z^{1\oplus u_s}\sigma_x^{u_{s'}}\}$ on the state $|\varphi\rangle$ ; $\sigma_z^{u_s}\sigma_x^{u_{s'}}|\varphi\rangle = \sigma_z^{1\oplus u_s}\sigma_x^{u_{s'}}|\varphi\rangle$ if $|\varphi\rangle \in \{|0\rangle, |1\rangle\}$. These operations by sender guarantee that receiver can get only either $\sigma_s \in \{I, \sigma_z\}$ or $\sigma_s \in \{\sigma_x, \sigma_z\sigma_x\}$ but not the exact Pauli operator. That is, receiver can successfully guess either sender has sent data $u_s u_{s'} \in \{00,10\}$ or $u_s u_{s'} \in \{01,11\}$ but not the definite 2-bit string $u_s u_{s'}$.

## Security analysis

In this section, we show that the power of two-fold quantum non-local correlations and special theory of relativity bounds both parties to remain fair and act according to the agreed codes: use genuine transformations, priorly agreed basis, and respond within allocated times. The underlying principle of unconditional security in the proposed framework is causality and two-fold quantum non-local correlations: first entanglement swapping and then teleportation over causally independent entangled systems. That is, for each $\sigma_a$ and every value of Alice's BSM result $\alpha\alpha'$, there will be unique swapped Bell state $|u_b u_{b'}\rangle$ and hence unique teleportation encoding $\sigma_i$ corresponding to BSM result of Bob $\beta\beta'$ as shown in table 1 & 2.

From $\sigma_b|\varphi\rangle$, $|\psi\rangle = \sigma_i\sigma_b|\varphi\rangle$, and measuring time lapse, Alice can verify that whether Bob's actions are consistent with his BSM result $\beta\beta'$ and corresponding teleportation operator $\sigma_i$ or not. On the other hand, if Alice's input $\sigma_a$ is consistent with Alice's BSM result $\alpha\alpha'$, swapped entangled state $|u_b u_{b'}\rangle$ (hence corresponding teleportation operator $\sigma_i$) and Alice replied within allocated time, Bob verifies that Alice is fair otherwise aborts the protocol.

In our proposed framework, security against mistrustful Alice and Bob lies in following requirements: (i) Alice (Bob) should not be able to know the definite Bob's input $\sigma_b$ (Alice's input $\sigma_a$) even at the end of protocol. (ii) Before or during the protocol, Alice (Bob) should not know the position $A_i$ ($B_i$) where Bob (Alice) will send the data until her (his) agents receive the data. Similarly, Alice (Bob) should not know the position of $B_i$ ($A_i$) even at the end of protocol. (iii) Proposed framework should not allow Alice (Bob) to alter her (his) input $\sigma_a$ ($\sigma_b$) from $\sigma_a \in \{I, \sigma_x\}$ to $\sigma_a \in \{\sigma_z, \sigma_z\sigma_x\}$ (from $\sigma_b \in \{I, \sigma_x\}$ to $\sigma_b \in \{\sigma_z, \sigma_z\sigma_x\}$) during the protocol and hence should allow Alice (Bob) to get joint measurement outcome $|\psi'\rangle = \sigma_a\sigma_i\sigma_b|\varphi\rangle$ of her (his) choice. (iv) Proposed framework should be resistant to Mayers and Lo-Chau attacks[41-43] where Alice can delay her input and try to influence measurement outcome $|\psi'\rangle = \sigma_a\sigma_i\sigma_b|\varphi\rangle$ by getting information about Bob's input $\sigma_b$. (v) Proposed framework should evade entanglement-based quantum attacks through non-local instantaneous computations. In other words, Alice responding from other than her announced position must be detected while Bob and B' must not be able to know Alice's input $\sigma_a$ during the protocol; before time $t = 2(x_a - x_b)$ where Alice reveals her oblivious input.

**(i) Data remains oblivious:** As for as Bob's data is concerned, Alice cannot find the exact value of $\sigma_b$ deterministically during or after the protocol – by measuring $\sigma_b|\varphi\rangle$, Alice can find that either $u_b u_{b'} \in \{00,01\}$ or $u_b u_{b'} \in \{10,11\}$ corresponding to Bob's operations $\sigma_b \in \{I, \sigma_x\}$ or $\sigma_b \in \{\sigma_z, \sigma_z\sigma_x\}$ but remains ignorant about the specific classical 2-bit string $u_b u_{b'}$ (or bit $b = u_b \oplus u_{b'}$) Bob has sent.

Similarly, Bob cannot differentiate between Alice's transformations $\sigma_z^{u_a}\sigma_x^{u_{a'}}$ or $\sigma_z^{u_a}\sigma_x^{1\oplus u_{a'}}$ on $|\psi\rangle$ since $\sigma_z^{u_a}\sigma_x^{u_{a'}}|\psi\rangle = \sigma_z^{u_a}\sigma_x^{1\oplus u_{a'}}|\psi\rangle$ where $|\psi\rangle = |\pm\rangle$. That is, if Bob gets $\psi' = \psi$ then he will be sure that Alice has applied $\sigma_a \in \{I, \sigma_x\}$ on $|\psi\rangle$ while $\sigma_a \in \{\sigma_z, \sigma_z\sigma_x\}$ in case of $\psi' \neq \psi$. As a result, Bob can find that either $u_a u_{a'} \in \{00,01\}$ or $u_a u_{a'} \in \{10,11\}$ corresponding to Alice's operations $\sigma_a \in \{I, \sigma_x\}$ or $\sigma_a \in \{\sigma_z, \sigma_z\sigma_x\}$ but remains ignorant about the specific classical 2-bit string $u_a u_{a'}$(or bit $a = u_a \oplus u_{a'}$) Alice has sent.

Since $\sigma_z^{u_a}\sigma_x^{u_{a'}}|\psi\rangle \neq \sigma_z^{u_a}\sigma_x^{1\oplus u_{a'}}|\psi\rangle$ if $|\psi\rangle \in \{|0\rangle, |1\rangle\}$, can Bob find exact operation by Alice on $|\psi\rangle$ by chossing different basis; computational one rather than agreed Hadamard basis? Interestingly, answer is NO. If B' deviates from agreed basis and sends $|\psi\rangle \in \{|0\rangle, |1\rangle\}$ to Alice, Alice will measure in Hadamard basis and get either $\psi = +$ or $\psi = -$ with equal probablity unknown to Bob. As a result, Bob will not be able to know whether Alice has applied tansformations $\sigma_a \in \{I, \sigma_x\}$ or $\sigma_a \in \{\sigma_z, \sigma_z\sigma_x\}$ on $|\psi\rangle$. By doing this, Bob will not get any advantages but rather he will allow Alice to cheat; since he cannot conclude now whether Alice's actions are consistant with non-local correlations or not.

**(ii) Transfer positions remain oblivious:** Before or during the protocol, Alice (Bob) cannot predict the position where Bob (Alice) will send the data. The choice of transfer position is totally random and Alice (Bob) can only know the transfer position $A_i$ ($B_i$) receives the data from Bob (Alice). Similarly, Alice (Bob) remains ignorant about transfer position $B_i$ ($A_i$) even after the protocol is complete - $B_i$ ($A_i$) do not communicate with Alice (Bob) during the protocol. Hence Alice (Bob) cannot compute time lapse or distance of the receiver $B_i$ ($A_i$).

**(iii) Zero biasness:** As we have stated earlier, underlying principle for unconditional security is quantum non-locality and causality; actions of space-like separated agents are causally independent and hence they cannot simulate their actions at some fixed time. Inputs $\sigma_a$ and $\sigma_b$ from Alice and Bob and their announced BSM results $\alpha\alpha'$ and $\beta\beta'$, result in unique swapped Bell state $|u_b u_{b'}\rangle$ and hence unique teleportation encoding $\sigma_i$. At the end of protocol, both parties know quantum states $\sigma_b|\varphi\rangle$ and $|\psi\rangle = \sigma_i\sigma_b|\varphi\rangle$. Hence both parties can validate or abort the protocol by looking whether the inputs $\sigma_a$ and $\sigma_b$ are consistent with non-local correlations among BSM results $\alpha\alpha'$ and $\beta\beta'$ and corresponding teleportation operator $\sigma_i$ or not.

Let's consider a simplest possible situation where $|u_a u_b\rangle = 00$, $|u_{a'} u_{b'}\rangle = 00$, $\sigma_a = I$ ($\sigma_a = \sigma_x$), and Alice announces BSM result $\alpha\alpha' = 00$ then $|u_b u_{b'}\rangle = 00$ ($|u_b u_{b'}\rangle = 01$). Now if BSM result of Bob is $\beta\beta' = 11$ while teleporting $\sigma_b|\varphi\rangle$ to B', then $\sigma_i = \sigma_z\sigma_x$ ($\sigma_i = \sigma_z$). In the revealing phase, if Alice acts fairly and replies $|\psi\rangle = I\sigma_i\sigma_b|\varphi\rangle$ ($|\psi\rangle = \sigma_x\sigma_i\sigma_b|\varphi\rangle$), Bob will verify that Alice's input $\sigma_a \in \{I, \sigma_x\}$ is consistent with teleportation encoding $\sigma_i = \sigma_z\sigma_x$ ($\sigma_i = \sigma_z$) by comparing $\sigma_b|\varphi\rangle$ and $|\psi\rangle = \sigma_i\sigma_b|\varphi\rangle$. However, if Alice tries to cheat by using different values of $\sigma_a = \sigma_z$ ($\sigma_a = \sigma_z\sigma_x$) and sends $|\psi\rangle = \sigma_z\sigma_i\sigma_b|\varphi\rangle$ ($|\psi\rangle = \sigma_z\sigma_x\sigma_i\sigma_b|\varphi\rangle$), Bob will extract $|u_b u_{b'}\rangle = 10$ ($|u_b u_{b'}\rangle = 11$) and hence different teleportation encoding $\sigma_i = \sigma_x$ ($\sigma_i = I$). As a result, Bob will abort the protocol by founding Alice's input $\sigma_a \in \{\sigma_z, \sigma_z\sigma_x\}$ inconsistent with the actual teleportation encoding $\sigma_i = \sigma_z\sigma_x$ ($\sigma_i = \sigma_z$). In conclusion Alice should not be able to change $\sigma_a$ from $\sigma_a \in \{I, \sigma_x\}$ to $\sigma_a \in \{\sigma_z, \sigma_z\sigma_x\}$ or $u_a u_{a'}$ from $u_a u_{a'} \in \{00, 01\}$ to $u_a u_{a'} \in \{10, 11\}$ after performing BSM on $\mathcal{H}_A \otimes \mathcal{H}_{A'}$ and she cannot do this in our proposed procedure.

Since $\mathcal{H}_S$ is also known to Alice, she can find swapped entangled state $|u_b u_{b'}\rangle$ corresponding to her input $\sigma_a$ and BSM result $\alpha\alpha'$. Hence she can validate or abort Bob's action by comparing $\sigma_b|\varphi\rangle$ (and $\beta\beta'$) and $|\psi\rangle = \sigma_i\sigma_b|\varphi\rangle$ received from Bob and B' respectively. Hence proposed framework guarantees that neither Alice nor Bob can influence measurement outcome of joint computation by altering their inputs during the protocol.

**(iv) Secure against MLC attacks:** Alice can choose Mayers and Lo-Chau attacks[41-43] and try to cheat as follow: Suppose she do not apply BSM on $\mathcal{H}_A \otimes \mathcal{H}_{A'}$, receives teleported state $|\psi\rangle = \sigma_i\sigma_b|\varphi\rangle$ from Bob over EPR channel $|u_a u_b\rangle$ and teleports another arbitrary state $|\varphi'\rangle$ to B' over EPR channel $|u_{a'} u_{b'}\rangle$ at $t = (x_a - x_b)$. She can know definite outcome of B' measurement instantly but cannot find $\sigma_b$ from $|\psi\rangle = \sigma_i\sigma_b|\varphi\rangle$ until she gets $\beta\beta'$ at $t = 2(x_a - x_b)$. Now in order to simulate her arbitrarily announced BSM result $\alpha\alpha'$ at $t = (x_a - x_b)$ with non-local correlations between $\sigma_b|\varphi\rangle$ and $|\psi'\rangle$, she has to apply specific $\sigma_a$ on $|\psi\rangle$ at $t = 2(x_a - x_b)$. Here Alice delayed input $\sigma_a$ can be either $\sigma_a \in \{I, \sigma_x\}$ or $\sigma_a \in \{\sigma_z, \sigma_z\sigma_x\}$ but she cannot give input of her choice at $t = 2(x_a - x_b)$ to get outcome she wants. In short, two-fold non-local quantum correlations and causality forces Alice to remain fair and perform agreed actions within time.

Similarly, Bob and B' cannot cheat through MLC attacks. Instead of sending a single qubit in the state $|\psi\rangle = |\pm\rangle$ at time $t = (x_a - x_b)$, suppose B' prepares an entangled quantum system $|\psi\rangle$ where

$$|\psi\rangle = \sum_i \lambda_i |\alpha_i\rangle |\beta_i\rangle \quad (3)$$

and sends system $|\alpha_i\rangle$ to Alice. Since he is causally disconnected from Bob, he cannot cheat by enfocing Alice to get valid non-local correlations by applying unitary transformations on $|\beta_i\rangle$.

**(v) Secure against attacks based on non-local computations:** Alice's input $\sigma_a$ can only be found before she reveals if someone has, Alice's BSM result $\alpha\alpha'$, Bob's BSM result $\beta\beta'$ and both quantum states $\sigma_b|\varphi\rangle$ and $|\psi\rangle = \sigma_i\sigma_b|\varphi\rangle$. Let's consider an example where $|u_a u_b\rangle = 00$, $|u_{a'} u_{b'}\rangle = 00$ and Bob and B' agree upon input $\sigma_b$ somewhere in causal past. Instantly at time $t = (x_a - x_b)$ when Bob teleports, eave (say B') can found $\sigma_i$ from $\sigma_b|\varphi\rangle$ and $|\psi\rangle = \sigma_i\sigma_b|\varphi\rangle$. Suppose eave founds $\sigma_i = \{\sigma_z, \sigma_z\sigma_x\}$ then he can extract, from table 2, that swapped entangled state would be $|u_b u_{b'}\rangle = \{00,01\}$ ($|u_b u_{b'}\rangle = \{10,11\}$) only if he knows that BSM result of Bob is $\beta\beta' = \{10,11\}$ ($\beta\beta' = \{00,01\}$). Now if eave extracts $|u_b u_{b'}\rangle = \{00,01\}$ say, table 1 shows that he can find Alice's input $\sigma_a = \{I, \sigma_x\}$ ($\sigma_a = \{\sigma_z, \sigma_z\sigma_x\}$) only if he knows her BSM result $\alpha\alpha' = \{00,01\}$ ($\alpha\alpha' = \{10,11\}$). Hence, if any piece of information from eave's set $E = \{\sigma_b|\varphi\rangle, \sigma_i\sigma_b|\varphi\rangle, \alpha\alpha', \beta\beta'\}$ is missing, Alice's input $\sigma_a$ cannot be extracted.

Suppose Bob and his agent B' have an arbitrary amount of pre-shared entanglement and have unlimited power of non-local instantaneous computations[46]. Even then they cannot find the complete set $E$ before time $t = 3(x_a - x_b)$. That is, they can only exchange classical information required to get outcome of non-local computations (Pauli encoding from $\sigma_b|\varphi\rangle$ and $|\psi\rangle = \sigma_i\sigma_b|\varphi\rangle$ here) at time $t = 3(x_a - x_b)$. Hence maximum commitment time is $2(x_a - x_b)$ where Alice commits at $t = (x_a - x_b)$, reveals at $t = 2(x_a - x_b)$, and Bob finds Alice's commitment at $t = 3(x_a - x_b)$ either fairly from $|\psi\rangle = \sigma_i\sigma_b|\varphi\rangle$ and $|\psi'\rangle = \sigma_a\sigma_i\sigma_b|\varphi\rangle$ or trying to extract from eave's set $E$. However, minimum commitment time would be $(x_a - x_b)$ if Bob assigns another agent at disjoint site nearby $(x_a, t_a)$ who can receive complete set $E$ at time $t = 2(x_a - x_b)$.

Similarly, Alice cannot cheat successfully by allowing her agents to perform non-local instantaneous computations (entanglement-based attacks) on her behalf. Suppose Alice assigns two more agents at $x_b < x < x_a$ and $x_a < x' < x_{b'}$ respectively. Since Bob teleports $\sigma_b|\varphi\rangle$ only at $t = (x_a - x_b)$, Alice's agents cannot get information about Bob's input $\sigma_b$ before Alice.

| $\|u_a u_b\rangle\|u_{a'} u_{b'}\rangle$ | | | | $(\alpha\alpha')\|u_b u_{b'}\rangle$ | | | |
|---|---|---|---|---|---|---|---|
| $\|00\rangle\ \|00\rangle$ | $\|01\rangle\ \|01\rangle$ | $\|10\rangle\ \|10\rangle$ | $\|11\rangle\ \|11\rangle$ | $(00)\ \|00\rangle$ | $(01)\ \|01\rangle$ | $(10)\ \|10\rangle$ | $(11)\ \|11\rangle$ |
| $\|00\rangle\ \|01\rangle$ | $\|01\rangle\ \|00\rangle$ | $\|10\rangle\ \|11\rangle$ | $\|11\rangle\ \|10\rangle$ | $(00)\ \|01\rangle$ | $(01)\ \|00\rangle$ | $(10)\ \|11\rangle$ | $(11)\ \|10\rangle$ |
| $\|00\rangle\ \|10\rangle$ | $\|01\rangle\ \|11\rangle$ | $\|10\rangle\ \|00\rangle$ | $\|11\rangle\ \|01\rangle$ | $(00)\ \|10\rangle$ | $(01)\ \|11\rangle$ | $(10)\ \|00\rangle$ | $(11)\ \|01\rangle$ |
| $\|00\rangle\ \|11\rangle$ | $\|01\rangle\ \|10\rangle$ | $\|10\rangle\ \|01\rangle$ | $\|11\rangle\ \|00\rangle$ | $(00)\ \|11\rangle$ | $(01)\ \|10\rangle$ | $(10)\ \|01\rangle$ | $(11)\ \|00\rangle$ |

**Table 1:** Entanglement swapping**:** This table shows all possible initial states of entangled particles $|u_a u_b\rangle|u_{a'} u_{b'}\rangle$ and corresponding outcomes of Alice's BSM $(\alpha\alpha')|u_b u_{b'}\rangle$. For example, if $|u_a u_b\rangle|u_{a'} u_{b'}\rangle = |00\rangle|11\rangle$, then swapped entangled pair $|u_b u_{b'}\rangle$ between Bob and B' will be in one of the four possible Bell sates: $|11\rangle$, $|10\rangle$, $|01\rangle$ and $|00\rangle$ corresponding to BSM result of Alice $\alpha\alpha'$ as 00, 01, 10, and 11 respectively.

| $\lvert u_b u_{b'}\rangle$ | Bob's BSM $\beta\beta'$ | | | | B' $\lvert\psi\rangle=\sigma_i\sigma_b\lvert\varphi\rangle$ | | | |
|---|---|---|---|---|---|---|---|---|
| $\lvert 00\rangle$ | 00 | 01 | 10 | 11 | $\sigma_b\lvert\varphi\rangle$ | $\sigma_x\sigma_b\lvert\varphi\rangle$ | $\sigma_z\sigma_b\lvert\varphi\rangle$ | $\sigma_z\sigma_x\sigma_b\lvert\varphi\rangle$ |
| $\lvert 01\rangle$ | 00 | 01 | 10 | 11 | $\sigma_x\sigma_b\lvert\varphi\rangle$ | $\sigma_b\lvert\varphi\rangle$ | $\sigma_z\sigma_x\sigma_b\lvert\varphi\rangle$ | $\sigma_z\sigma_b\lvert\varphi\rangle$ |
| $\lvert 10\rangle$ | 00 | 01 | 10 | 11 | $\sigma_z\sigma_b\lvert\varphi\rangle$ | $\sigma_z\sigma_x\sigma_b\lvert\varphi\rangle$ | $\sigma_b\lvert\varphi\rangle$ | $\sigma_x\sigma_b\lvert\varphi\rangle$ |
| $\lvert 11\rangle$ | 00 | 01 | 10 | 11 | $\sigma_z\sigma_x\sigma_b\lvert\varphi\rangle$ | $\sigma_z\sigma_b\lvert\varphi\rangle$ | $\sigma_x\sigma_b\lvert\varphi\rangle$ | $\sigma_b\lvert\varphi\rangle$ |

**Table 2:** Teleportation: This table shows all possible Bell states $\lvert u_b u_{b'}\rangle$ swapped between Bob and B' due to BSM of Alice, Bob's BSM results $\beta\beta'$ on his part of the entangled pair and state $\sigma_b\lvert\varphi\rangle$ and corresponding possibilities of state $\lvert\psi\rangle$ on the B' side. For example, if Bob and B' have share entangled state as $\lvert 01\rangle$ and BSM result of Bob is $\beta\beta'=10$ then B' will have state $\lvert\psi\rangle=\sigma_z\sigma_x\sigma_b\lvert\varphi\rangle$ on his side.

In conclusion, the proposed framework is secure for the purpose of OT, two-sided TPSC and unbiased ideal coin tossing since Alice/Bob cannot influence the final outcome $f(\sigma_a,\sigma_b;\lvert\varphi\rangle)=\sigma_a\sigma_i\sigma_b\lvert\varphi\rangle$ of her/his choice. Similarly, it fulfills the security requirements for bit commitment where committer Alice is not allowed to change her commitment while receiver Bob remains unable to know the commitment before it is revealed by Alice.

## Discussion

We proposed a general relativistic quantum framework for cryptography based on non-local quantum correlations and theory of relativity. The framework determines the actions of both parties through causality and quantum non-locality. Quantum non-local correlations assure that communicating parties act fairly while impossibility of superluminal signaling is used for insuring timely responses.

The framework is based on interesting combination of non-locality and theory of special relativity that gives new notion of OT; the receiver can only get specific information about the data but not its exact identity. That is, receiver may know both the transferred messages but remains oblivious about the genuine one. Moreover, the transfer position remains oblivious to the sender throughout the protocol while receiver can find the exact position only when he/she receives the data. The sender is guaranteed that the receiver can gain specific information about the data that logically follows from the protocol and know the transfer position only if the protocol is completed and the receiver acts fairly. Moreover, if the receiver completes the protocol successfully, he will be certain that the transferred data has come from the legitimate sender. The authenticity and integrity of the data transferred is guaranteed, the receiver rejects the data if the sender tries to modify it after the protocol has been started.

This fascinating combination of EPR type quantum correlations with causal structure of Minkowski space time shows the power of relativistic quantum cryptography in defining tasks that are considered to be impossible in non-relativistic cryptography. For example, unconditionally secure and deterministic two-sided TPSC, asynchronous ideal coin tossing with zero bias and unconditionally secure bit commitment.

Although it is standard in mistrustful quantum cryptography that both parties have efficient quantum technologies (quantum computer), the proposed relativistic quantum framework can be reliably implemented without requiring quantum computer. Both parties can calculate $f(\sigma_a, \sigma_b; |\varphi\rangle)$ securely with existing quantum technologies; photo detectors without needing long term quantum memory. However, even having quantum computers, neither party can cheat successfully.

The proposed framework can easily be generalized for other multiparty mistrustful cryptographic tasks securely. For example, if B' is considered to be a third party instead of Bob's agent then our proposed framework could be used to implement secure quantum secret sharing between sender Bob and receivers Alice and B'. The secret $\sigma_b$ can be divided among Alice and B' in terms of classical BSM result $\alpha\alpha'$ and quantum state $|\psi\rangle = \sigma_a \sigma_i \sigma_b |\varphi\rangle$ respectively.

We hope this work would open new directions in quantum information, quantum computation, quantum cryptography and their connections with special theory of relativity. On the other hand, proposed framework is purely relativistic quantum mechanical where both input and output data is associated with unitary transformations applied on quantum systems and it does not require any secure classical channels; classical information can be publically announced. Hence, it would in return prove to be helpful in developing our understanding about the true description of the world, the quantum theory.